\documentstyle[multicol,epsf,bm,prl,aps]{revtex}

\begin{document}

\draft
\preprint{Preprint No.}
\title{Stimulated emission with a non-equilibrium state of radiation}
\author{L.Accardi\thanks{accardi@volterra.mat.uniroma2.it}, ~K.Imafuku\thanks{imafuku@volterra.mat.uniroma2.it}, ~S.V.Kozyrev$^1$\thanks{kozyrev@mi.ras.ru}}
\address{Centro Vito Volterra, Universita' di Roma Tor Vergata, 
00133 Rome, Italy\\
N.N. Semenov Institute of Chemical Physics, Russian Academy of Science, \\
117334 Moscow, Russia$^1$}
\date{April 24, 2001}
\maketitle
\begin{abstract}
The stimulated emission from an atom interacting with radiation 
in non-equilibrium state is considered. 
The stochastic limit,  applied to the  
non-relativistic Hamiltonian describing the interaction, shows 
that the state of atoms,  driven by some non-equilibrium 
state of the field approaches a stationary state which can 
continuously emit photon, unlike the case 
with an equilibrium state.
\end{abstract}

Einstein applied Planck's radiation theory to describe the equilibrium 
state between a atom and field\cite{ref:Einstein}.  He perceived that such an 
equilibrium state can be realized through the detailed balance 
condition, i.e. the balance in each mode between spontaneous and 
``stimulated emission", i.e. the emission from the atom 
stimulated by the field. Einstein's investigation can be said to give 
the clearest insight into the origin of Planck's radiation because, with this 
notion, we can understand the Planck's law on the density of states of 
the photons from an equilibrium field, i.e.
\begin{equation}
\rho(\omega)d\omega=\frac{\hbar \omega^3}{\pi^2 c^3}\frac{1}
{\exp\left(\frac{\hbar\omega}{kT}\right)-1}d\omega .
\end{equation}

On the other hand, with the development of technology,
a controlled  emission with a controlled stimulating 
field has been realized in experimental situation,
for example, a laser system.
A laser is often described in terms of stimulated emission 
from an equilibrium state at {\it negative temperature} \cite{ref:laser}. 
However  the concept of the negative temperature can't be accepted literally, 
but rather as an expediency to understand an inverse 
population state of the atoms which can emit photons.
More precisely, a laser should be considered as a stimulated emission 
due to a non-equilibrium state of a field.
In this letter we propose a way to understand such kind of
stimulated emissions from non-equilibrium states, 
without introducing any phenomenological expediency like 
a negative temperature. 

We apply the stochastic limit \cite{ref:Accardi_text} 
to the non-relativistic QED Hamiltonian and derive two types of 
equations: one is the so-called rate equation for the atom and the other 
is  an equation describing the time evolution of the number of photons.  
In addition, based on these equations, 
we discuss the stimulated emission from the atom interacting 
with some non-equilibrium field.  The state of an atom driven by the 
non-equilibrium state of the field approaches 
a stationary state which can continuously emit photon, unlike the 
case of the equilibrium field.  The conditions of  
the non-equilibrium state  of the field for such phenomena 
is also made clear.

We shall consider an atom interacting with the EM-filed 
described with the standard non-relativistic QED Hamiltonian
\begin{equation}
H=H_0+\lambda ~V,\quad H_0=H_A+H_F
\end{equation}
where
\begin{equation}
H_A=\frac{p^2}{2m}+V(q),\quad
\left([q,p]=i\right)
\end{equation}
\begin{equation}
H_F=\int \omega(k) a^{\dagger}_{k,\sigma}a_{k,\sigma},~\quad 
\left([a_{k,\sigma},a^{\dagger}_{k',\sigma'}]=\delta_{\sigma\sigma'}\delta(k-k')\right)
\end{equation}
and
\begin{equation}
V= \sum_{\sigma} \int dk \frac{1}{|k|^{1/2}}
\varepsilon_{k,\sigma}\left(a_{\sigma,k}e^{ik\cdot q}
+a_{\sigma,k}^\dagger e^{-ik\cdot q}\right).
\end{equation}
$\sigma$ is polarization index ($\sigma=\leftrightarrow,\updownarrow$).
In the following discussion we assume that $H_A$ has discrete 
spectrum or 
$
H_A|\epsilon_a\rangle=\epsilon_a|\epsilon_a\rangle
$. 
The interaction Hamiltonian in interaction picture 
$V(t)$ can be witten as \cite{ref:Accardi_Kozyrev_note}
\begin{equation}
V(t)=\sum_{\omega \in F} \sum_{a,b}\int dk \left(
\overline{g_{ab}(k,\sigma)}
E_\omega^{\dagger}\Big(|\epsilon_a \rangle\langle \epsilon_b|\Big)
a_{k,\sigma} e^{-i(\omega(k)-\omega)t}
+h.c \right)
\end{equation}
where
\begin{eqnarray}
\overline{g_{ab}(k,\sigma)}
&=&\frac{1}{|k|^{1/2}} 
~\langle \epsilon_a|e^{ik\cdot q}p \cdot \varepsilon_\sigma|\epsilon_b\rangle,
\\
E^\dagger_\omega(X)&=&
\sum_{\epsilon_r \in F_\omega}P_{\epsilon_r}
XP_{\epsilon_r-\omega},\quad 
P_{\epsilon_r}:=|\epsilon_r\rangle\langle \epsilon_r|\\
F&=&\{\omega=\epsilon_r-\epsilon_r';~\epsilon_r,\epsilon_r'\in Spec ~H_A\},\quad\mbox{(Bohr frequency)}\\
F_\omega&=&\{\epsilon_r\in Spec ~H_s ;~\epsilon_r-\omega\in Spec ~H_A\}.
\end{eqnarray}
We investigate the dynamics of the system described by the above 
Hamiltonian using the stochastic limit, which describes the quantum 
dynamics in the regime of weak coupling ($\lambda\rightarrow 0$) 
and large times ($t \rightarrow t/\lambda^2$) \cite{ref:Accardi_text}.  
The main result of this theory (stochastic golden rule) 
\cite{ref:Accardi_text} is that the time rescaling $t\rightarrow t/\lambda^2$ 
induces a rescaling of the quantum field
\begin{equation}\label{rescaled_fielsd}
a_{k,\sigma}\rightarrow\frac{1}{\lambda}e^{-i\frac{t}{\lambda^2}(\omega(k)-\omega)}a_{k,\sigma}
\end{equation}
and, in the limit $\lambda\rightarrow 0$ the rescaled field 
(\ref{rescaled_fielsd}) becomes a quantum white noise (or master field)
$b_{\omega,\sigma}(t,k)$ satisfying the commutation 
relations
\begin{equation}
[b_{\omega,\sigma}(t,k),b_{\omega',\sigma'}^\dagger(t',k')]
=\delta_{\sigma\sigma'}\delta_{\omega\omega'}2\pi \delta(t-t')\delta(\omega(k)-\omega)\delta(k-k').
\end{equation}
Moreover, if the initial state of the field is 
the mean zero gauge invariant Gaussian state with correlations 
$$\langle a^\dagger_k a_k'\rangle=N(k)\delta(k-k')$$, 
then the state of the limit white noise will be of the some 
type with correlations
\begin{eqnarray}\label{correlation1}
\langle b_{\omega,\sigma}(t,k)
b_{\omega',\sigma'}^\dagger (t',k') \rangle &=&
\delta_{\sigma\sigma'}\delta_{\omega\omega'}2\pi \delta(t-t')\delta(\omega(k)-\omega)\delta(k-k')N_{\sigma}(k)\\
\langle b^\dagger_{\omega,\sigma}(t,k)
b_{\omega',\sigma'}(t',k') \rangle &=&
\delta_{\sigma\sigma'}\delta_{\omega\omega'}2\pi \delta(t-t')\delta(\omega(k)-\omega)\delta(k-k')(N_{\sigma}(k)+1).\label{correlation2}
\end{eqnarray}
The Schr\"{o}dinger equation becomes a quantum white noise equation 
which, after having been normally ordered, takes the form 
\cite{ref:Accardi_text,ref:Accardi_Kozyrev_note}
\begin{equation}
dU_t=(-idH(t)-Gdt)U_t\quad;\quad t>0
\end{equation}
with the initial condition $U_0=1$ and where
$dH(t)$, {\it called the martingale term}, is the stochastic differential:
\begin{equation}
dH(t)=\sum_{\omega \in F} \sum_{a,b}\int dk \left(
\overline{g_{ab}(k,\sigma)}
E_\omega^{\dagger}\Big(|\epsilon_a \rangle\langle \epsilon_b|\Big)
dB_{\omega,\sigma}(t)+
E_\omega\Big(|\epsilon_a \rangle\langle \epsilon_b|\Big)
dB_{\omega,\sigma}^\dagger(t)
\right)
\end{equation}
driven by the quantum Brownian motions
\begin{equation}
dB_{\omega,\sigma}=\sum_{\epsilon_a-\epsilon_b=\omega}\int^{t+dt}_{t}d\tau
~\int dk \overline{g_{ab}(k,\sigma)}b_{\omega,\sigma}(\tau,k),
\end{equation}
and the operator $G$, {\it called the drift term}, is given by
\begin{equation}
G=\sum_{\sigma}\sum_{\omega \in F} \left(
(g|g)^{-}_{\omega,\sigma}
E^\dagger_{\omega}\Big(|\epsilon_a \rangle\langle \epsilon_b|\Big)
E_{\omega}\Big(|\epsilon_a \rangle\langle \epsilon_b|\Big)
+\overline{(g|g)}^{+}_{\omega,\sigma}
E_{\omega}\Big(|\epsilon_a \rangle\langle \epsilon_b|\Big)
E^\dagger_{\omega}\Big(|\epsilon_a \rangle\langle \epsilon_b|\Big)
\right)
\end{equation}
\begin{eqnarray}
(g|g)^{-}_{\omega,\sigma}&=&\sum_{\epsilon_a-\epsilon_b=\omega}\int dk |g_{ab}(k,\sigma)|^2 \frac{-i(N_{\sigma}(k)+1)}{\omega(k)-\omega-i0},\label{19}\\
(g|g)^{+}_{\omega,\sigma}&=&\sum_{\epsilon_a-\epsilon_b=\omega}\int dk |g_{ab}(k,\sigma)|^2 \frac{-iN_{\sigma}(k)}{\omega(k)-\omega-i0}.\label{20}
\end{eqnarray}
With this quantum stochastic differential equation, we can always 
derive equations for the atom and field, that is, 
(i)master equation for reduced density operator for the atom,
and 
(ii)equation for the field\cite{ref:Accardi_Kozyrev_note,ref:Accardi_KI_Kozyrev}.

One should notice that we do not assume the state of the field to be 
equilibrium. Therefore $N_{\sigma}(k)$ in (\ref{correlation1}) 
and (\ref{correlation2}) is not necessary to be 
$$
N_{\sigma}(k)=\frac{1}{e^{\beta \omega(k)}-1}, 
~\left(\beta=\frac{1}{kT}\right),
$$
but can be, for some general nonlinear functions 
$\beta_{\sigma}(\omega)$,
\begin{equation}\label{nesf}
N_{\sigma}(k)=\frac{1}{e^{\beta_{\sigma}\left(\omega(k)\right)}-1}.
\end{equation}

In this letter, let us consider a 3-level atom 
($\epsilon_1 < \epsilon_2 < \epsilon_3$) whose matrix 
elements satisfy the conditions (Fig.~1)
\begin{equation}\label{assumption0}
\langle \epsilon_1| e^{ikq}p \cdot\varepsilon_{\updownarrow}
|\epsilon_2\rangle=0,
\quad
\langle \epsilon_2| e^{ikq}p \cdot
\varepsilon_{\leftrightarrow}|\epsilon_3\rangle
=\langle \epsilon_1| e^{ikq}p \cdot
\varepsilon_{\leftrightarrow}|\epsilon_3\rangle=0,\quad
others\neq 0. 
\end{equation}
There are rather standard conditions in laser theory~\cite{ref:laser}.
Notice however that condition (\ref{assumption0}) requires the 
preparation of the atom in a situation in which the longitudinal
($\updownarrow$) and transverse ($\leftrightarrow$) polarization 
do not enter symmmetrically. We believe that atoms, satisfying the condition 
(\ref{assumption0}), can be experimentally prepared.
In addition, for simplicity, we restrict ourselves to a generic 
system \cite{ref:Accardi_Kozyrev_note}: this means in our case that the 
3 Bohr frequencies  $\omega_{21}$, $\omega_{31}$, $\omega_{32}$  
($\omega_{jk}=\epsilon_{j}-\epsilon_{k}$) are all different among 
themselves.

With these assumptions, we can derive the so-called rate equation 
for the atom 
\begin{eqnarray}\label{rate_3_eq2_1}
\frac{d}{dt}P_1(t)&=& -\left\{
2(\gamma_{21,\leftrightarrow}^{(+)}+\gamma_{31,\updownarrow}^{(+)})P_{1}(t)
-2(\gamma^{(-)}_{21,\leftrightarrow}P_{2}(t)+\gamma_{31,\updownarrow}^{(-)}
P_{3}(t))
\right\}\nonumber\\
\frac{d}{dt}P_2(t)&=& -\left\{
2(\gamma_{21,\leftrightarrow}^{(-)}+\gamma_{32,\updownarrow}^{(+)})P_{2}(t)
-2(\gamma^{(-)}_{32,\updownarrow}P_{3}(t)+\gamma_{21,\leftrightarrow}^{(+)}P_{1}(t))
\right\}\nonumber\\
\frac{d}{dt}P_3(t)&=& -\left\{
2(\gamma_{32,\updownarrow}^{(-)}+\gamma_{31,\updownarrow}^{(-)})P_{3}(t)
-2(\gamma^{(+)}_{31,\updownarrow}
P_{1}(t)+\gamma_{32,\updownarrow}^{(+)}P_{2}(t))
\right\}\nonumber\\
\label{rate_equation}
\end{eqnarray}
and the corresponding equations for  polarized photons
\begin{eqnarray}\label{n_tr}
\frac{d}{dt}n_{\leftrightarrow}(t)&=&
2~\left(\gamma^{(-)}_{21,\leftrightarrow}P_2(t)
-\gamma^{(+)}_{21,\leftrightarrow}P_1(t)\right)\label{n_lr(t)}\\
\frac{d}{dt}n_{\updownarrow}(t)&=&2~\left(\gamma^{(-)}_{31,\updownarrow}P_3(t)
-\gamma^{(+)}_{31,\updownarrow}P_1(t)+
\gamma^{(-)}_{32,\updownarrow}P_3(t)
-\gamma^{(+)}_{32,\updownarrow}P_2(t)\right)\label{n_ud(t)}
\end{eqnarray}
where
\begin{equation}
P_j(t)=Tr\Big(\rho_{tot}(0)U^\dagger_t|\epsilon_j\rangle\langle\epsilon_j|U_t \Big),\quad
n_{\sigma}(t)=Tr\Big(\rho_{tot}(0)U^\dagger_t \int dk 
a^\dagger_{k,\sigma}a_{k,\sigma}~U_t \Big),
\end{equation}
\begin{equation}
\gamma_{ij,\sigma}^{(\pm)}={\rm Re}(g|g)^{\pm}_{\omega_{ij},\sigma},
\end{equation}
and the $(g|g)^{\pm}_{\omega_{ij},\sigma}$ 
are given by (\ref{19}) and (\ref{20}).
Equations (\ref{rate_equation}) have a stationary solution 
which satisfies 
\begin{eqnarray}\label{st_sol_2}
\frac{P_2(\infty)}{P_1(\infty)}&=&
\frac{\gamma_{31,\updownarrow}^{(-)}\gamma_{21,\leftrightarrow}^{(+)}+
\gamma_{32,\updownarrow}^{(-)}\gamma_{21,\leftrightarrow}^{(+)}+
\gamma_{31,\updownarrow}^{(+)}\gamma_{32,\updownarrow}^{(-)}}
{\gamma_{31,\updownarrow}^{(-)}\gamma_{21,\leftrightarrow}^{(-)}+
\gamma_{31,\updownarrow}^{(-)}\gamma_{32,\updownarrow}^{(+)}+
\gamma_{32,\updownarrow}^{(-)}\gamma_{21,\leftrightarrow}^{(-)}}
\nonumber\\
\frac{P_3(\infty)}{P_1(\infty)}&=&
\frac{
\gamma_{31,\updownarrow}^{(+)}\gamma_{21,\leftrightarrow}^{(-)}+
\gamma_{32,\updownarrow}^{(+)}\gamma_{21,\leftrightarrow}^{(+)}+
\gamma_{31,\updownarrow}^{(+)}\gamma_{32,\updownarrow}^{(+)}}
{\gamma_{32,\updownarrow}^{(-)}\gamma_{21,\leftrightarrow}^{(-)}+
\gamma_{31,\updownarrow}^{(-)}\gamma_{21,\leftrightarrow}^{(-)}+
\gamma_{31,\updownarrow}^{(-)}\gamma_{32,\updownarrow}^{(+)}}
\nonumber\\
\frac{P_3(\infty)}{P_2(\infty)}&=&
\frac
{\gamma_{32,\updownarrow}^{(+)}\gamma_{21,\leftrightarrow}^{(+)}+
\gamma_{31,\updownarrow}^{(+)}\gamma_{21,\leftrightarrow}^{(-)}+
\gamma_{32,\updownarrow}^{(+)}\gamma_{31,\updownarrow}^{(+)}}
{\gamma_{31,\updownarrow}^{(-)}\gamma_{21,\leftrightarrow}^{(+)}+
\gamma_{32,\updownarrow}^{(-)}\gamma_{21,\leftrightarrow}^{(+)}+
\gamma_{31,\updownarrow}^{(+)}\gamma_{32,\updownarrow}^{(-)}}.
\nonumber\\
\end{eqnarray}

The right-hand side of equation (\ref{n_tr}) shows that, 
if
$
\gamma^{(-)}_{21,\leftrightarrow}P_2(t)
-\gamma^{(+)}_{21,\leftrightarrow}P_1(t) > 0
$,
i.e. if
\begin{equation}
\frac{P_2(t)}{P_1(t)} > \frac{\gamma_{21,\leftrightarrow}^{(+)}}{\gamma_{21,\leftrightarrow}^{(-)}}
\end{equation}
then, $n_{\leftrightarrow}(t)$ must increase.  
{\it Also  with the  stationary state of the atom}, 
$n_{\leftrightarrow}(t)$ must keep increasing if
\begin{equation}\label{emission_condition}
\frac{P_2(\infty)}{P_1(\infty)} > \frac{\gamma_{21,\leftrightarrow}^{(+)}}{\gamma_{21,\leftrightarrow}^{(-)}}
\end{equation}
is satisfied. This means that, if we can experimentally realize 
condition (\ref{emission_condition}), then we can also realize a 
continuous emission of 
$\leftrightarrow$-photon from the atom, stimulated by 
a non-equilibrium state of the field. 
Using (\ref{st_sol_2}), the condition (\ref{emission_condition}) 
can be written as 
\begin{equation}\label{condition_0}
\frac{\gamma_{21,\leftrightarrow}^{(+)}}{\gamma_{21,\leftrightarrow}^{(-)}}
< \frac{\gamma_{31,\updownarrow}^{(-)}\gamma_{21,\leftrightarrow}^{(+)}+
\gamma_{32,\updownarrow}^{(-)}\gamma_{21,\leftrightarrow}^{(+)}+
\gamma_{31,\updownarrow}^{(+)}\gamma_{32,\updownarrow}^{(-)}}
{\gamma_{31,\updownarrow}^{(-)}\gamma_{21,\leftrightarrow}^{(-)}+
\gamma_{31,\updownarrow}^{(-)}\gamma_{32,\updownarrow}^{(+)}+
\gamma_{32,\updownarrow}^{(-)}\gamma_{21,\leftrightarrow}^{(-)}}
~~\Leftrightarrow
~~\frac{\gamma_{21,\leftrightarrow}^{(+)}}{\gamma_{21,\leftrightarrow}^{(-)}}
<
\frac{\gamma_{31,\updownarrow}^{(+)}}{\gamma_{31,\updownarrow}^{(-)}}
\frac{\gamma_{32,\updownarrow}^{(-)}}{\gamma_{32,\updownarrow}^{(+)}}.
\end{equation}
If we suppose that the photon densities in the state (\ref{nesf}) satisfy
$$
N_{\leftrightarrow}(\omega)=N_{\updownarrow}(\omega)=\frac{1}{e^{\beta(\omega)}-1},
$$
the condition (\ref{condition_0}) becomes equivalent to
\begin{equation}\label{condition_1}
\beta(\omega_{31}) < \beta(\omega_{32})+\beta(\omega_{21}).
\end{equation}
This process does not break the energy-conservation law. In fact 
one can easily check that 
\begin{equation}
\frac{d}{dt} \left(\sum_{j=1}^3 \epsilon_jP_j(t)+\sum_{\sigma}
\int dk~\omega(k) Tr\Big(\rho_{tot}(0)U^\dagger_t 
a^\dagger_{k,\sigma}a_{k,\sigma}~U_t \Big) \right)=0,
\end{equation}
and one should notice that 
the energy of the $\updownarrow$-field would be 
converted to the energy of the $\leftrightarrow$-field through the 
stationary state of the atom (Fig.~2).
This is  a peculiar property of the 
stimulated emission from a  non-equilibrium state of the field. 
In fact the usual Gibbs states are characterized by the fact 
that the function $\beta$ in (\ref{nesf}) is linear, i.e. 
$\beta(\omega)=\beta \omega$, and in this case condition (\ref{condition_1}) is never satisfied.
The non-equilibrium field can therefore drive the atom to a stationary 
state which can emit photon, and this means that this stationary state 
gives an example of dissipative structure in the Prigogine 
sense~\cite{ref:Prigogine}.

In order to understand better the physical meanings of our condition 
(\ref{assumption0}), 
let us consider an extreme case such as  
\begin{equation}
N_{\leftrightarrow}(\omega) \ll N_{\updownarrow}(\omega).
\end{equation}
In this situation, 
the relation (\ref{st_sol_2}), defining the stationary solution, 
becomes approximately 
\begin{eqnarray}
\frac{P_2}{P_1}&\sim& \frac{\gamma_{31,\updownarrow}^{(+)}
\gamma_{32,\updownarrow}^{(-)}}
{\gamma_{32,\updownarrow}^{(-)}\gamma_{31,\updownarrow}^{(+)}}
=\frac{N_{\updownarrow}(\omega_{32})}{N_{\updownarrow}
(\omega_{32})+1}\frac{N_{\updownarrow}(\omega_{31})+1}
{N_{\updownarrow}(\omega_{31})}
\left(\neq \frac{N_{\updownarrow}(\omega_{21})}
{N_{\updownarrow}(\omega_{21})+1},~\mbox{in general but Gibbs}\right)
\label{double_einstein}\\
\frac{P_3}{P_1}&\sim& \frac{\gamma_{31,\updownarrow}^{(+)}}
{\gamma_{31,\updownarrow}^{(-)}}\qquad=
\frac{N_{\updownarrow}(\omega_{31})}{N_{\updownarrow}
(\omega_{31})+1}\\
\frac{P_3}{P_2}&\sim& \frac{\gamma_{32,\updownarrow}^{(+)}}
{\gamma_{32,\updownarrow}^{(-)}}\qquad=\frac{N_{\updownarrow}(\omega_{32})}
{N_{\updownarrow}(\omega_{32})+1}.
\end{eqnarray}
Notice that (\ref{double_einstein}) is the {\it Double Einstein formula}
discussed in \cite{ref:Accardi_KI_Kozyrev} and that 
this stationary state is not a detailed balance but a {\it distorted balance} 
state. 
It is obvious that we need at least a 3-level atom to get such 
a distorted balance state because, since 
detailed balance means balance at each transition frequency, 
it follows that, if there is only one such frequency 
as in a 2-level atom, then every stationary 
state is a detailed balance state.
This distorted balance state should play the role of the {\it negative temperature} state introduced in the phenomenology of laser systems\cite{ref:laser}.
Under the assumption (\ref{assumption0}) $\updownarrow$-field can be 
interpreted as the {\it pumping field} which realizes the so-called 
inverse population state of atom and the  
$\leftrightarrow$-field  as a {\it stimulating field} which stimulates 
the emission from 2 to 1.

Finally, let us discuss to the stability of the non-equilibrium 
state of the field.  
With the above analysis, the non-equilibrium state seems
extremely fragile against the interaction with atoms, while the 
equilibrium state seems robust. 
This difference should play an essential 
role to understand the dynamical properties of the 
nonequilibrium state of the field itself.
It is an interesting problem to consider whether the non-equilibrium 
state after some interaction can become stable or not, and 
this problem should be related to the discussion on the stability 
of the macroscopic state. 
In order to approach these problems, we would have to take account of some kind of self-interaction of the field, that is the re-stimulation by 
the emitted photon, which we neglected in our discussion.
Taking into account these effects, the equations of this system 
would be non-linear equations between $P_i(t)$, $n_{\sigma}(t)$ 
and other quantities, unlike in this letter.  
Our treatment in this letter is not sufficient to deal with 
the stability problem of the field.
But in another application of the stochastic limit to 
superfluidity\cite{ref:Accardi_Kozyrev:superfluid}, 
a procedure to deal with this kind of self interaction 
has been already 
discussed and a non-linear equation has been derived. 
We think that we will be able to approach  the stability problems of 
the non-equilibrium field also with a similar procedure.

\bigskip
The authors are grateful to I.V.Volovich for discussion.
Kentaro Imafuku and Sergei Kozyrev are grateful to Centro Vito 
Volterra and Luigi Accardi for kind hospitality.
This work was partially supported by INTAS 9900545 grant. 
Kentaro Imafuku is supported by a overseas research fellowship 
of Japan Science and Technology Corporation.
Sergei Kozyrev was partially supported by RFFI 990100866 grant.



\newpage
\vspace*{2cm}

\begin{figure}
\epsfxsize=0.9\textwidth
\begin{center}
\epsfbox{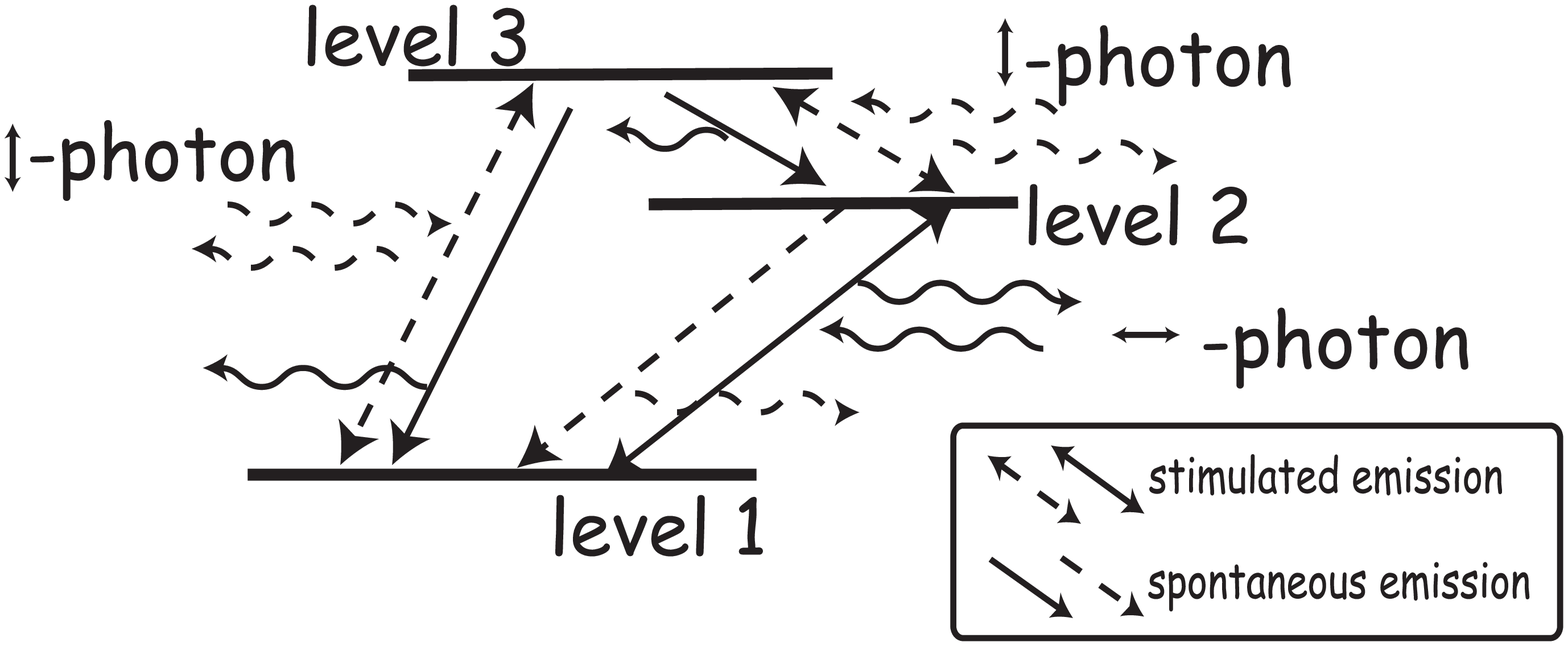}
\end{center}
\caption{Schematical illustration of our assumption (\ref{assumption0}).}
\label{fig1}

\vspace{2cm}

\epsfxsize=0.9\textwidth
\begin{center}
\epsfbox{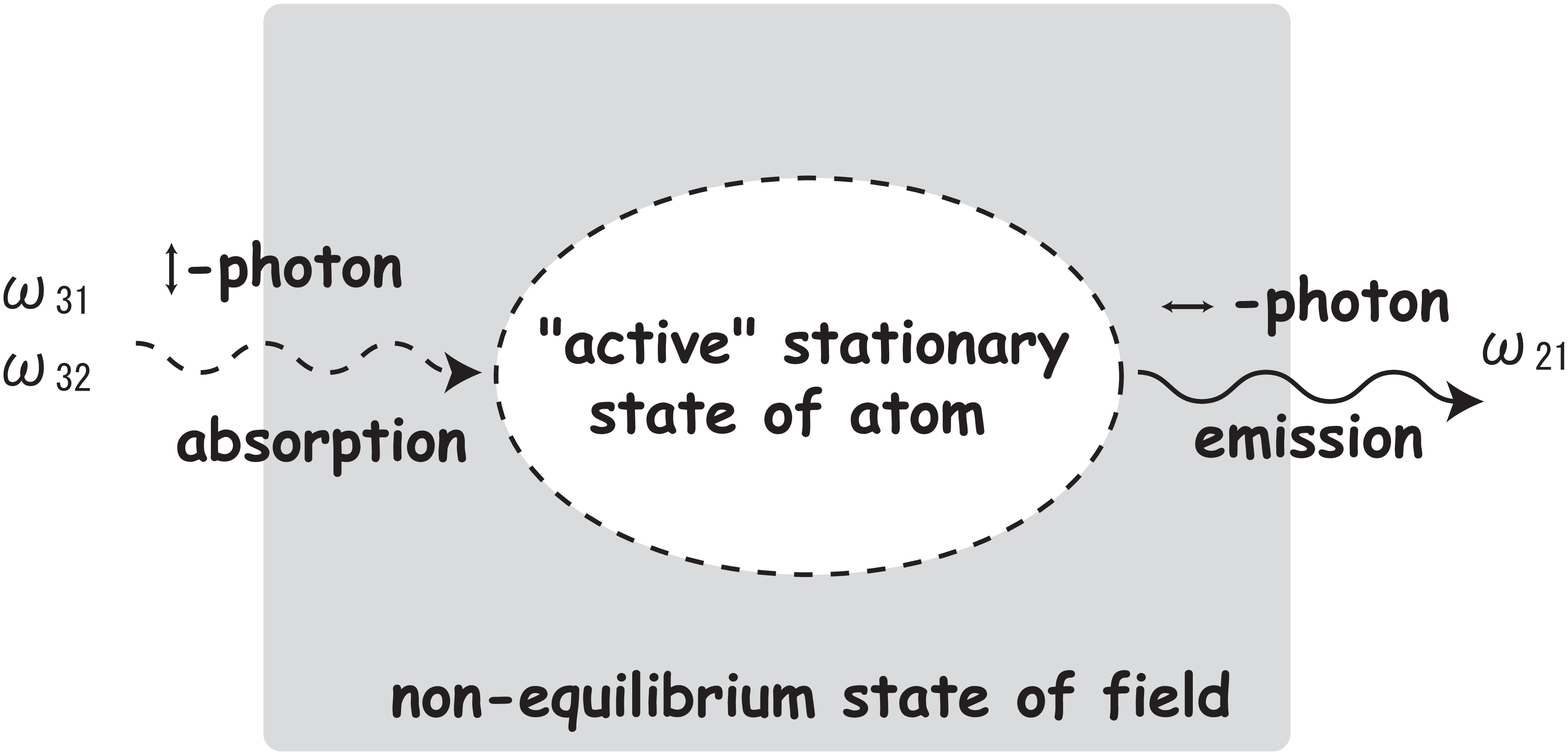}
\end{center}
\caption{The energy flow from $\updownarrow$-photon to 
$\leftrightarrow$-photon.}
\label{fig2}
\end{figure}

\end{document}